\documentclass[%
reprint,
 amsmath,amssymb,
aps,
]{revtex4-2}
\usepackage{multirow}
\usepackage{graphicx}
\usepackage{dcolumn}
\usepackage{bm}
 
\begin{document}

\preprint{APS/123-QED}

\title{Room-temperature quasi-continuous-wave pentacene maser pumped by an invasive Ce:YAG luminescent concentrator}

\author{Hao Wu}

\author{Xiangyu Xie}
\author{Wern Ng}
\author{Seif Mehanna}
\author{Yingxu Li}
\author{Max Attwood}
\author{Mark Oxborrow}
 \email{m.oxborrow@imperial.ac.uk}
\affiliation{%
 Department of Materials, Imperial College London, Exhibition Road, London, SW7 2AZ, UK
}%

\date{\today}

\begin{abstract}
We present in this  work a quasi-continuous-wave (CW) pentacene maser operating at 1.45 GHz in the  Earth's magnetic field at room temperature with a duration of $\sim$4 ms and an output power of up to -25 dBm. The maser is optically pumped by a cerium-doped YAG (Ce:YAG) luminescent concentrator (LC) whose wedge-shaped output is embedded inside a 0.1\% pentacene-doped \textit{para}-terphenyl (Pc:Ptp) crystal. The pumped crystal is located inside a ring of strontium titanate (STO) that supports a TE$_{01\delta}$ mode of high magnetic Purcell factor. Combined with simulations, our results indicate that CW operation of pentacene masers at room-temperature is perfectly feasible so long as excessive heating of the crystal is avoided. 
\end{abstract}

\maketitle


\section{Introduction}

The maser\cite{gordon1955maser}, the microwave analogue of a laser, is a device that exploits stimulated emission to  amplify electromagnetic signals at GHz frequencies with very little added noise. By offering
extreme sensitivity (\textit{i.e.}~noise temperatures of just a few Kelvin)\cite{culver1957maser}, exceptionally low intermodulation distortion \cite{schulz-dubois64}, and rock-solid durability (in the face of jamming or accidental input overload), solid-state masers\cite{chang1958characteristics,siegmann1964microwave} have proven themselves useful for deep-space communication links\cite{shell1994ruby,clauss08}. When configured as an oscillator, both gaseous\cite{gordon1954molecular} and solid-state masers can be exploited as secondary frequency standards for time keeping and metrology\cite{vanier1982active,bourgeois2005maser}.
It was noted early on \cite{bloembergen60} that zero-field masers, though not magnetically tunable,
dispense with the need for carefully grown mono-crystals of known cut and orientation.

In the case of solid-state gain media such as ruby (based on covalently or ionically bonded crystals), cryogenic conditions are required for maser action. Operation at room temperature is precluded by the extremely rapid increase in the rate of spin-lattice relaxation with absolute temperature. In 2012, an optically-pumped solid-state maser operating at 1.45~GHz, based on pentacene-doped \textit{para}-terphenyl (Pc:Ptp), was demonstrated\cite{oxborrow2012room}, though maser oscillation occurred in bursts lasting only $\sim$250~$\mu$s. This device exploited emissive spin polarization (\textit{i.e.} a population inversion) across the $|\textrm{X}\rangle$ and $|\textrm{Z}\rangle$ sublevels of pentacene's lowest lying triplet state, T$_1$, reached from pentacene's ground state, S$_0$, via photo-excitation at $\lambda=590$ nm (from S$_0$ to S$_1$), followed by sublevel (\textit{i.e.} spin)-selective intersystem crossing (from S$_1$ to T$_2$), followed by internal conversion (from T$_2$ to T$_1$). Despite providing a high polarization density and thus a high instantaneous output power (of around $-10$~dBm), it was suspected that the build-up of population in T$_1$'s lowerst and longest lived sublevel, $|\textrm{Z}\rangle$, destroys the initial population inversion. Analogous to the workings of a nitrogen laser, this ``bottlenecking''\cite{oxborrow2017maser} in the flow of population should cause the maser to self-terminate, as was indeed observed\cite{salvadori2017nanosecond,breeze2015enhanced}. Alternative/superior gain media were thus sought\cite{jin2015proposal,bogatko2016molecular,van2017candidate, fischer2018highly}. A breakthrough came with the demonstration of a room-temperature maser employing charged nitrogen-vacancy (NV$^-$) defects in diamond\cite{breeze2018continuous} that was capable of CW operation. This demonstration required a uniform applied magnetic field of $\sim430$ mT and the output power of the maser (configured as a oscillator) was vastly lower ($-90$~dBm). As was pointed out in ref.~\cite{oxborrow2017maser}, to provide low-noise amplification at room temperature, the maser's co-operativity ($C$ in ref.~[\onlinecite{breeze2018continuous}] and this work) must vastly exceed unity. In other words, the active electromagnetic gain provided by the maser's pumped medium (exhibiting an extremely small negative spin temperature) must substantially outstrip the passive electromagnetic losses incurred by dissipation within the room-temperature materials (\textit{viz.} dielectric and metals) that compose the maser's electromagnetic cavity. Since there has been no report yet of an
NV$^-$ diamond maser operating in the high-\textit{C} regime, the potential of such masers to amplify more quietly
(\textit{i.e.} to demonstrate lower noise temperatures) than conventional microwave amplifiers incorporating high-electron-mobility transistors (HEMTs)
remains unproven. The high doping concentrations (parts per thousand) and thus high spin-polarization densities attainable with Pc:Ptp (together with usefully if not spectacularly long decay/decoherence times) offers (i) hope of accessing the elusive  ``$C \gg 1$'' regime needed for low-noise amplification at room temperature,
as well as (ii) a route to the realization of  strong-coupling effects\cite{breeze2017room}.

 Recently, we accurately investigated the zero-field spin dynamics of the (lowest lying) triplet state of pentacene when doped in \textit{para}-terphenyl  at room-temperature\cite{wu2019unraveling} and found the dynamics to lie tantalizingly close to the boundary between the ``bottlenecked'' and ``CW-masing-feasible" regimes. Encouraging results with pump pulses of longer duration motivated us to re-examine the feasibility of CW operation.
In this paper, we present a  room-temperature pentacene maser working in a quasi-continuous wave mode (quasi-CW) with an output power up to -25 dBm and a duration of $\sim$4 ms. This duration was as long as that of the optical pump pulse applied to the gain medium. We attribute this accomplishment to three key modifications over the original pentacene maser: i) the use of a different optical pump source, ii) improved optical coupling of the pump light into the maser crystal and iii) the use of a maser cavity supporting a microwave mode of high magnetic Purcell factor. 

Unlike conventional optical pumping with pulsed lasers\cite{oxborrow2012room,salvadori2017nanosecond}, we use  a cerium-doped YAG luminescent concentrator (Ce:YAG LC) pumped by a xenon flash lamp. This arrangement provides intense pulses of yellow light that are both well-matched spectrally to Pc:Ptp's optical absorption bands and can be substantially longer in duration (\textit{i.e.} milliseconds) than what a ``long''-pulsed dye laser is capable of\cite{sathian2017solid}. Though high-power solid-state CW (and CW \textit{Q}-switched) lasers outputting yellow light are now available, they are still relatively complex and bulky machines with mediocre wall-plug efficiencies (a few \%). We point out that high (mono-mode) beam quality and diffraction-limited optics, enabling (sub-)milliradian beam divergences and micron spot sizes, are not needed for illuminating millimetre-sized maser crystals. 

Compared to non-laser light sources, such as lamps and LEDs, LCs can (when pumped by the former two) provide light at higher output intensities and/or radiances, resulting in greater pump irradiances at the target compared to direct illumination. This is because -as the name itself suggests- LCs advantageously \textit{concentrate} photonic energy/power: primary photons (in our case coming from a xenon flash lamp) that enter the LC through large input faces are converted (``red-shifted'') to lower-frequency photons, which then exit through the LC's far smaller output surface(s). Though there is a price (energetically) of red-shifting, combined with various other loss mechanisms (such as self-absorption) that further reduce the overall optical power, the concentration process ultimately leads to a substantial increase in the number of photons traversing an aperture per unit time, per unit area of aperture (per steradian of direction), hence boosting the light's intensity and/or radiance\cite{de2016high,barbet2016light}. High-efficiency InGaN LEDs, outputting blue light, can very effectively pump Ce:YAG\cite{sathian2017solid}. Though spectrally less efficient, a xenon flash lamp can provide the required primary pump light (across the wavelengths Ce:YAG usefully absorbs) at far greater instantaneous intensity/radiance than what a bank of InGaN LEDs can. The use of such a flash lamp can thus enable the generation of higher instantaneous intensity/radiance at the LC's output, thus giving an increased capacity for optical pumping of the maser crystal to and beyond the threshold for masing.

Furthermore, the output end of our LC is wedge-shaped and embedded into the Pc:Ptp crystal, implementing what we refer to as ``invasive optical pumping''\cite{wu2020invasive}. Compared to simply butt-coupling the LC's end face (cut flat and at right-angles to the LC's optical axis) onto a flat external face of the maser crystal, this wedged geometry (i) reduces the retro-reflection of luminescence photons back into the LC, (ii) guides light from the LC more deeply into the maser crystal, thus distributing the illumination (\textit{i.e.} ``pumping'') more uniformly throughout the crystal's bulk, (iii) provides a hermetic seal (excluding atmospheric oxygen) at the interface where the pump light enters the Pc:Ptp crystal, and (iv) provides an effective thermal anchorage of the Pc:Ptp crystal to a solid ``substrate'' material (namely Ce:YAG) exhibiting high thermal conductivity and volumetric heat capacity. Attributes (ii) to (iv) help to avoid thermal and/or photo-chemical damage (scorching) at the crystal's surface. With regard to (i), modifying the shape of the LC's output surface(s) from a flat end to that of a wedged end\cite{oxborrow2017maser} has been proven to enhance the efficiency at which light is extracted from the LC\cite{gallinelli2019enhancing}. With regard to the gradual splaying out of the directions of light (to angles of incidence below the critical angle, thus allowing escape) that is achieved with a taper or wedge, a single ``chisel'' shaped wedge is as effective as a symmetrical ``vee''\cite{oxborrow2017maser}, provided the wedging angle is small. For simplicity (ease of making), we here adopt the former with respect to the LC's thickness direction.

Moreover, a miniaturized maser cavity made of a strontium titanate (STO) crystal was employed to boost the spin-photon coupling strength (associated with~Einstein's $B$ co-efficient) via the Purcell effect\cite{purcell1946spontaneous}, thus lowering the optical pumping threshold\cite{breeze2015enhanced}. By simulating the obtained 4-ms maser burst obtained experimentally, we found that pentacene's underlying spin dynamics supports CW maser operation at room-temperature, provided the heat dissipated within the maser crystal by optical pumping can be removed. 
 \section{Principle of maser operation -- triplet mechanism}
\begin{figure}[htbp!]
\centering\includegraphics{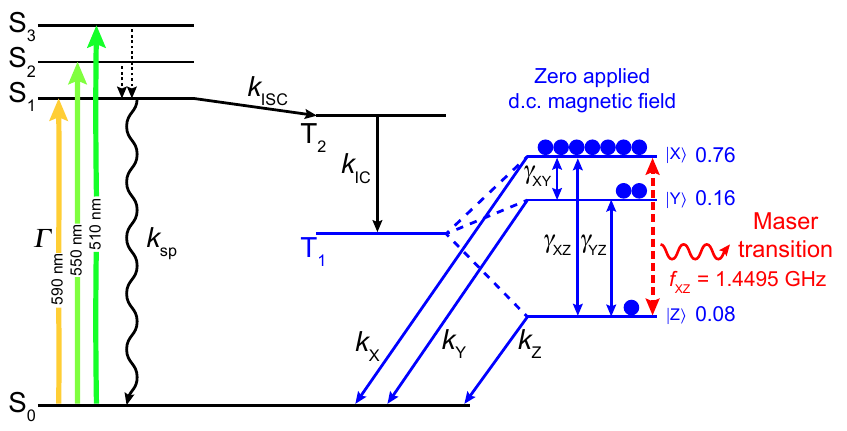}
\caption{Jablonski diagram showing the optical pumping scheme that prepares highly polarized pentacene triplets (blue) at zero applied magnetic field. The maser transition with a frequency of 1.4495 GHz occurs between the $|\textrm{X}\rangle$ and $|\textrm{Z}\rangle$ triplet sublevels. $\mathit{\Gamma}$: pump rate; $k_\textrm{sp}$: spontaneous emission rate; $k_\textrm{ISC}$: intersystem crossing rate; $k_\textrm{IC}$: internal conversion rate; $\gamma_\textrm{XY}$: spin-lattice relaxation rate between sublevels $|\text{X}\rangle$ and $|\text{Y}\rangle$, similarly for $\gamma_\textrm{XZ}$  and $\gamma_\textrm{YZ}$; $k_\textrm{X}$: the decay rate from sublevel $|\text{X}\rangle$ back to the ground state S$_0$, similarly for $k_\textrm{Y}$ and $k_\textrm{Z}$.}
\label{fig.1}
\end{figure}

 The pentacene maser is pumped through the triplet mechanism (``TM'') as shown in Fig.~\ref{fig.1}.
Individual pentacene molecules substitutively doped into the \textit{para}-terphenyl crystal are optically pumped from the singlet ground state S$_0$ to the singlet excited state S$_i$ ($i = 1,2,3,...$) for which the pump wavelengths of 590 nm, 550 nm and 510 nm have been determined to be optimal for inducing $\textrm{S}_0\rightarrow \textrm{S}_1$, $\textrm{S}_0\rightarrow \textrm{S}_2$ and $\textrm{S}_0\rightarrow \textrm{S}_3$ transitions respectively\cite{bogatko2016molecular}. Optically excited pentacene molecules can transition from S$_1$ back to S$_0$ via either stimulated or spontaneous optical emission, the rate of the former growing with the pump light's intensity\cite{takeda2002zero}. Since it can be ignored in this work, the majority ($\sim$ 62.5\%\cite{takeda2002zero}) of the pentacene molecules in S$_1$ will resonantly transfer to the triplet state T$_2$ by spin-orbit coupling mediated by intersystem crossing (ISC). The ISC process is highly spin-selective, resulting in non-Boltzmann distributions of populations across the triplet state's sublevels, $|\text{X}\rangle$, $|\text{Y}\rangle$ and $|\text{Z}\rangle$ in the experimentally determined ratio of $0.76:0.16:0.08$\cite{sloop1981electron}. Pentacene molecules in T$_2$ rapidly decay via internal conversion (IC) to the lowest triplet state T$_1$ while preserving the population distribution (and hence spin polarization) across T$_1$'s sublevels. This provides the prerequisite condition for  masing: a strong initial population inversion across $|\text{X}\rangle$ and $|\text{Z}\rangle$. When the highly polarized pentacene molecules are placed in a cavity supporting a microwave mode at 1.4495 GHz, which matches the zero-field splitting between $|\text{X}\rangle$ and $|\text{Z}\rangle$ \cite{breeze2015enhanced}, stimulated emission is induced by microwave photons occupying the mode. Alongside stimulated emission, spin-lattice relaxation and non-radiative decay of triplets back to S$_0$ affect the populations of the three sublevels of T$_1$. Masing occurs when the gain associated with stimulated emission (minus the loss associated with stimulated absorption across the same $|\text{X}\rangle \leftrightarrow|\text{Z}\rangle$ transition) exceeds the dissipative decay rate for 1.45 GHz photons in the mode. 
\section{Maser assembly}
\begin{figure*}[htbp!]
\centering\includegraphics{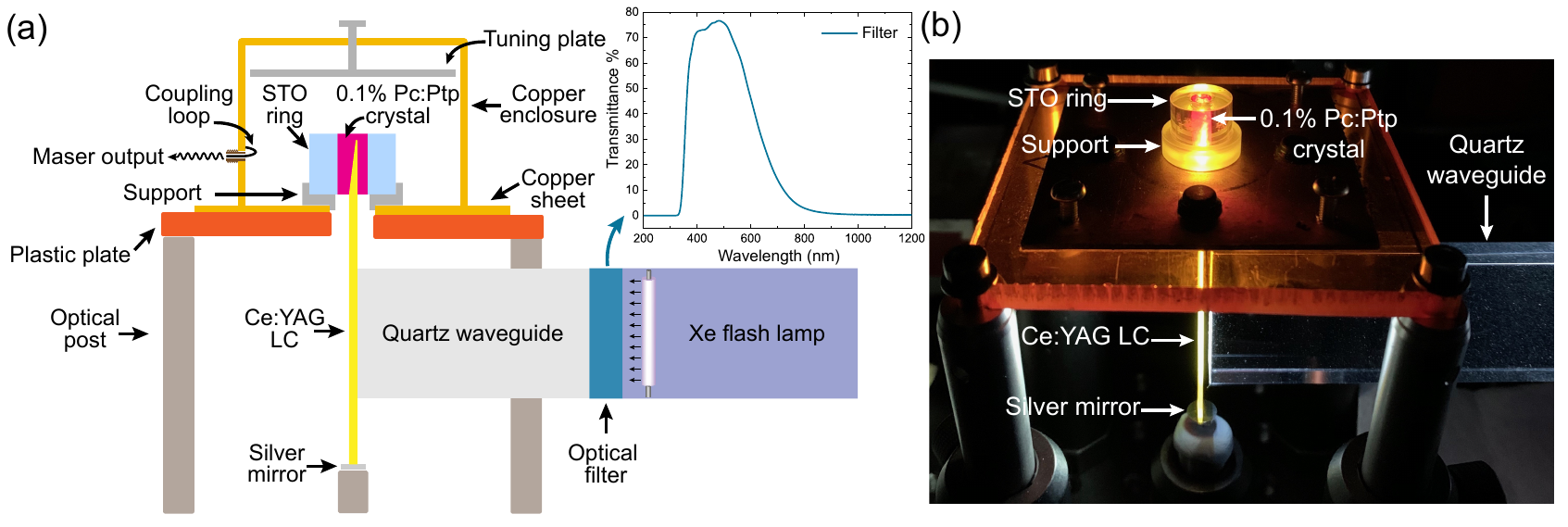}
\caption{Setup of the Ce:YAG LC-pumped pentacene maser. (a) A schematic diagram of the resonator and optical setup. Inset: UV/vis spectrum of a filter installed at the output of the xenon flash lamp. (b) Photograph of the setup (without showing the copper enclosure of the STO cavity, the optical filter and the xenon flash lamp). } \label{fig.2}
\end{figure*}

The experimental setup of the quasi-CW pentacene maser is shown in Fig.~\ref{fig.2} with all critical components labelled. Salient details about the maser assembly are described below.
\subsection{Optical pumping}
\begin{figure}[b!]
\centering\includegraphics{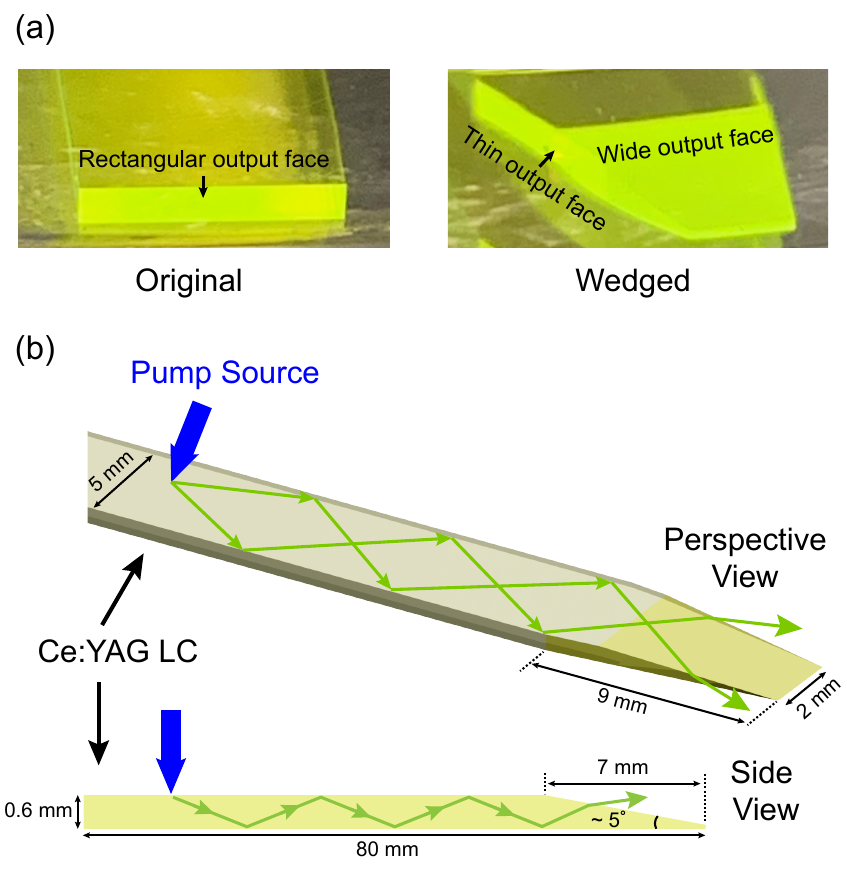}
\caption{Concept and design of the invasive optical pumping with the wedged Ce:YAG LC. (a) Photograph highlighting the original and wedged output faces of the Ce:YAG LC under room light. (b) Schematic of the wedged Ce:YAG LC with its dimensions. With optical pumping, the induced luminescence (green lines) escapes the LC from its wedged output faces.} \label{fig.3}
\end{figure}

A luminescent concentrator (LC) was received from Crytur \cite{crytur17} in the form of a slender bar (rectangular prism) of ``Cryphosphor''-branded Ce:YAG; its dimensions were 80 mm long ($l$) $\times$ 5 mm wide ($w$) $\times$ 0.6 mm thick ($d$)
with all six of its faces polished. Additional faces were then flat-ground and polished by hand onto one end of the bar. These faces comprised (i) one wide isosceles-trapezoidal face reducing the bar's thickness and (ii) a symmetrical pair of narrow triangular faces reducing the bar's width. The latter allowed the wedged end to fit into the 4-mm bore of the maser cavity's STO dielectric ring (see Fig.~\ref{fig.3}).
The LC's output faces, injecting light into the crystal, comprised the above three faces plus the section of the bar's untouched $l \times w$ face lying opposite the trapezoidal face (i).
To maximise the output at the wedged end, the $w \times d$ face at the opposite end of the Ce:YAG bar was
attached to a small solid-silver mirror, which reflected light back into the bar towards its wedged end.

The Ce:YAG LC was pumped by a xenon lamp housed within the water-cooled output head of a Lumenis Vasculight intense-pulsed-light (``IPL'') source designed for aesthetic skin treatments. Light from the head was conveyed to the LC via a  polished fused-silica  waveguide (supplied by Robson Scientific; 150 mm long $\times$ 34 mm wide $\times$ 8 mm thick); its outputting end face was stationed $\sim$~2~mm from one of the LC's 5-mm-wide faces. A glass band-pass filter transmitting wavelengths between 350 nm and 800 nm (inset of Fig.~\ref{fig.2}(a)) was installed between the glass flow tube (containing the xenon lamp) of the IPL's head and the waveguide's input face. This guaranteed that no harmful ultraviolet light could escape from the IPL head,
and reduced the unnecessary heating of the LC by infrared light from the lamp's hot xenon plasma. 

\begin{figure}
\centering\includegraphics{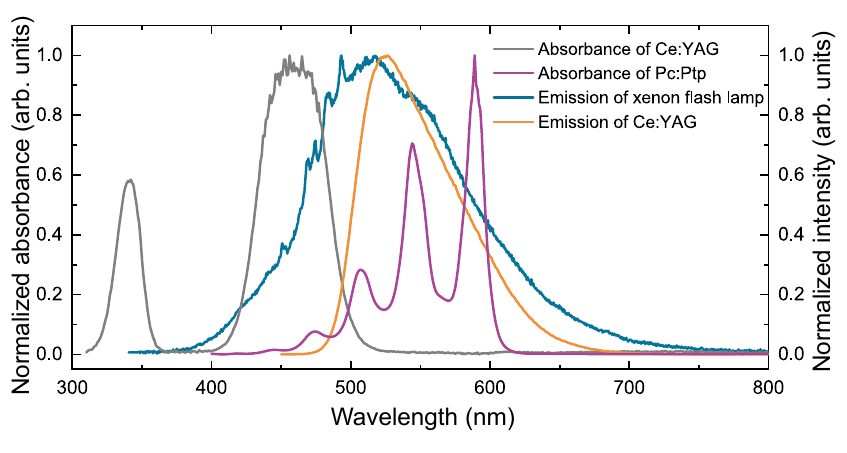}
\caption{UV/Vis spectra of Ce:YAG (grey) and Pc:Ptp (purple) and emission spectra of Ce:YAG (orange) and xenon flash lamp (cyan).}
\label{fig.4}
\end{figure}

Optical characterisations in terms of the absorptive and emissive properties of the xenon flash lamp, Ce:YAG LC and 0.1\% Pc:Ptp crystal were carried out to verify the feasibility of the optical pumping scheme. Optical absorption spectra of the Ce:YAG LC and Pc:Ptp crystal were measured by an Agilent Cary 5000 UV-Vis-NIR spectrophotometer. The photoluminescence (PL) of the Ce:YAG LC was measured by a Horiba Fluorolog-3 spectrofluorometer, exciting at  410 nm (this choice of 
excitation wavelength avoided saturating the instrument's sensitive photo-multiplier detector). 
The emission spectrum of the xenon flash lamp was measured by coupling light from the fused-silica waveguide's output face into the input aperture of a home-made integrating sphere\cite{sathian2017solid}; light from the output aperture of the sphere was fed into an  Ocean Optics USB2000+ spectrometer. 
The emission and absorption spectra obtained from the above measurements are shown together in Fig.~\ref{fig.4}. With the glass band-pass filter in place, the output of the xenon flash lamp shows a broad peak ranging from 350 nm to 800 nm indicating the effective elimination of ultraviolet and infrared light together with a good spectral match to Ce:YAG's visible absorption band in the blue, peaking at $\sim$~460 nm. The measured emission from the Ce:YAG luminescent concentrator ranges from 480 nm to 700 nm and overlaps with the transition wavelengths of $\textrm{S}_0\rightarrow \textrm{S}_1$ (590 nm), $\textrm{S}_0\rightarrow \textrm{S}_2$  (540 nm) and $\textrm{S}_0\rightarrow \textrm{S}_3$  (510 nm) for pentacene molecules within the Pc:Ptp crystal, which implies that a Ce:YAG LC should work as an effective, albeit imperfect,  pump source for pentacene masers. It is worth noting that although the emission spectrum of the xenon flash lamp also overlaps with the pentacene's absorption bands, the LC's prismatic geometry and specular surfaces, as shown
in Figs.~\ref{fig.2}~and~\ref{fig.3}, substantially eliminate any direct pumping of the Pc:Ptp crystal by the xenon flash lamp, ensuring that all the light absorbed by pentacene molecules within the Ptp crystal derives from the Ce:YAG's luminescence.

\subsection{Gain medium}
\begin{figure}
\centering\includegraphics{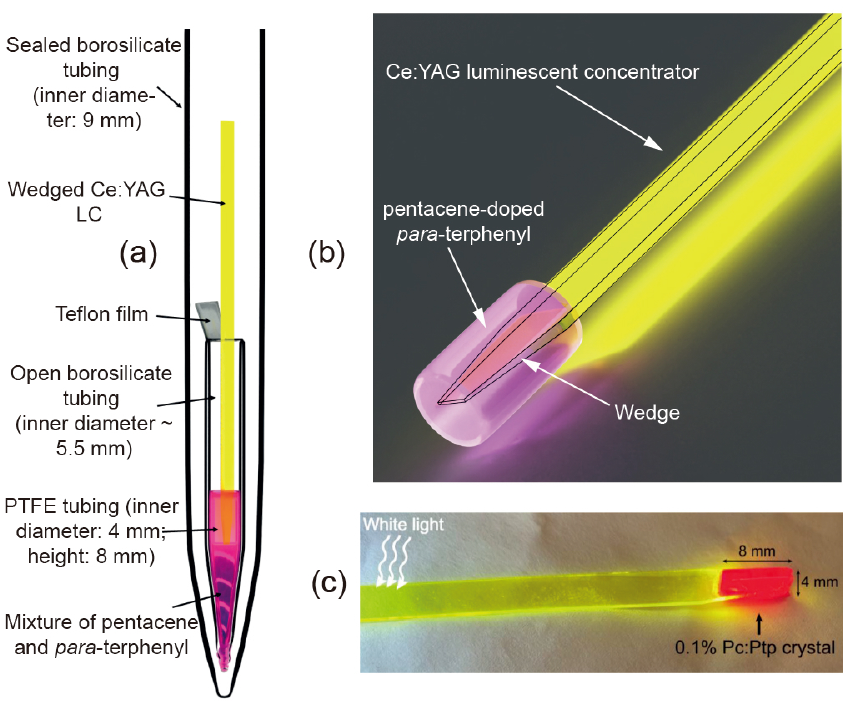}
\caption{(a) 3D-rendered image of the crystal-growth assembly (the upper, empty section of the outer glass vial is not shown). (b) Computer-generated image of the cylindrical Pc:Ptp maser crystal (pink) grown around the wedged end of the Ce:YAG LC (lemon yellow) (c) Actual photograph of the red-fluorescing 0.1\% Pc:Ptp crystal pumped with yellow light from the Ce:YAG (lemon yellow), which itself is pumped by pseudo-white light from a mobile phone's torch.} \label{fig.5}
\end{figure}

The maser gain medium was a \textit{para}-terphenyl (Ptp) crystal substitutionally doped with pentacene (Pc) molecules
at a molar ratio of 0.1\% (nominal). Using the vertical Bridgman-Stockbarger method, this crystal was grown around the wedged end of the Ce:YAG bar to form a ``lollipop'', with the wedge
functioning as an internal mould (see Figs.~\ref{fig.5}(b) \& (c)). Pentacene (TCI Europe NV) was used as received and \textit{para}-terphenyl (Sigma--Aldrich, $\geq$99.5\%) was purified by zone-refining with 30 cycles. Both were finely ground into powder and mixed in the molar
ratio of $1:1000$. The mixture was loaded into a small clean borosilicate tube (inner diameter (I.D.) $\sim$5.5 mm). A cylindrical poly-tetrafluoroethylene (PTFE) sleeve with an inner diameter of 4 mm and a height of 8 mm was fitted around the wedged end of the LC. This sleeve acted as an external mold limiting the crystal's radial growth to the same diameter. 
Several bent layers of PTFE film  were gently wedged between the LC and the lip of the small borosilicate tube; these  prevented the LC from sliding downwards during the crystal's growth. The tube containing the mixed powder, the LC and the PTFE sleeve was lowered into a larger borosilicate tube (Dixon Science,  9 mm I.D., already flame-sealed at its bottom end) which was then flame-sealed at its top end under an argon atmosphere (BOC, 99.9999\%, $\sim$800~mbar pressure). The entire assembly is shown in Fig.~\ref{fig.5}(a). The sealed outer tube was dangled by a nichrome wire from the end of a cantilever attached to the translation stage of a home-constructed vertical translator. It was then lowered into a vertical tube furnace (Elite Thermal Systems Ltd.) with the tapered end of the smaller internal borosilicate tube  positioned at the furnace's central (hottest) zone, set to a temperature of 217 $^{\circ}$C (just sufficient to melt the mixed powder). 
Using the vertical translator, the vial was lowered towards a cooler part of the tube furnace
at a rate of 2 mm per hour. Once it was certain that the whole  melt had re-solidified (by knowing the furnace's vertical temperature profile), the oven's temperature was ramped downwards at the rate of $\sim10\ ^{\circ}$C per hour.  
A nominally 0.1\% Pc:Ptp crystal was thereby formed around  the wedged end of the Ce:YAG LC after a 3-day growth and cooling run. Cracking away the outer and inner borosilicate tubes, then carefully peeling away the inner PTFE sleeve, left a pink cylindrical bead of Pc:Ptp crystal on a prismatic, lemon-yellow shank of Ce:YAG. The wedged output faces of the Ce:YAG luminescent concentrator were entombed within the bead; the resulting ``lollipop'' is shown in Fig.~\ref{fig.5}(b) and (c). The bead of Pc:Ptp fitted snugly into the STO ring shown in Fig.~\ref{fig.2}.

By growing the Pc:Ptp crystal on the wedged end of the LC, the LC's guided luminescence can be ``injected''
efficiently into the bulk of the crystal; this is the aforementioned ``invasive optical pumping'' approach\cite{wu2020invasive}.
This approach is uniquely available to optically-pumped masers since, in contrast to a laser crystal, the optical homogeneity of the gain medium does not matter. Optical-supply waveguides (acting like ``arteries'' for light flow)
of a different refractive index to the maser gain medium itself (such as the wedge inserted \textit{within} the Pc:Ptp bead) can advantageously (w.r.t.~pumping) invade the gain medium without adversely affecting the ``beam quality'' of the electromagnetic wave being amplified. The only downside of using these supply arteries is that the space they occupy could otherwise be filled with gain medium; in other words, the maser crystal's filling factor w.r.t.~the microwave mode will be reduced. 
The refractive index of Pc:Ptp is 1.61\cite{de1979fluorescence} and that of 
Ce:YAG is $1.83$\cite{zelmon1998refractive}; their combination, with the latter located inside the former, forms
an optical waveguide where reflections off the wedged faces gradually increase the angle of incidence (for pump photons in the waveguide's core impinging on its cladding) until the critical angle is exceeded and the pump photons escape into the Pc:Ptp around it. 

Previous arrangements have in contrast used conventional external (``side'') optical pumping where,
due to the large mismatch between the refractive indices of Ce:YAG and that of air ($=1 $),
an optical coupling medium (typically oil or glue) is placed between the LC's output face and the receiving face of the gain medium\cite{barbet2016light,sathian2017solid}. But such an arrangement is problematic at high optical pump fluences due to photochemical damage/decomposition of either the coupling fluid or the  Pc:Ptp crystal itself in the presence of oxygen. The charring of either can readily lead to photo-thermal run-away. 
Here, in contrast, the LC's output faces are directly in contact with the Pc:Ptp crystal, forming a hermetic seal excluding oxygen.
The invasive Ce:YAG wedge also functions as a thermal substrate (``heat sink'') for the Pc:Ptp, helping to keep it cool.
These two features of invasive optical pumping help to protect the Pc:Ptp from charring. The bright red fluorescence emanating from the Pc:Ptp crystal shown in Fig.~\ref{fig.5}(c)
demonstrates the efficient optical coupling of the LC's output into the crystal.

\subsection{Maser cavity}
\begin{figure}
\centering\includegraphics{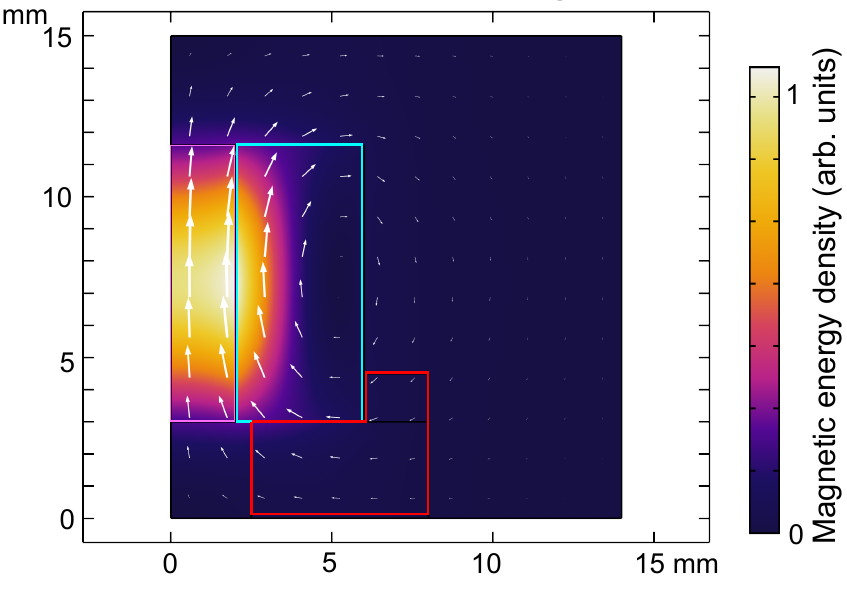}
\caption{2D axisymmetric simulation of the magnetic energy density (false-color ``heat'' map) and magnetic field vector (white arrows) of the TE$_{01\delta}$ mode of the  STO-loaded microwave  cavity. The geometries of the STO ring, Pc:Ptp crystal and cross-linked polystyrene support in medial cross-section  are highlighted in blue, magenta and red, respectively. The width of the entire region of the simulation (14 mm) corresponds to the radius of the copper enclosure of the resonator, while the height of the region (15 mm) corresponds to the height between the tuning plate and the copper sheet at the bottom of the resonator.} \label{fig.6}
\end{figure}
The maser cavity used here was an STO ring (Gaskell Quartz Ltd.,
I.D.~=~4.05~mm,
outer diameter (O.D.)~=~12.0 mm, height~=~8.6~mm) housed inside a cylindrical copper enclosure. A support made of cross-linked polystyrene (viz.~Polypenco~Q200.5 Elder Engineering Ltd.) raised the STO ring
3~mm above a copper conducting plane. Fig.~\ref{fig.2} shows the cavity setup with the Pc:Ptp crystal loaded inside the STO ring. An inductive loop (coupling coefficient $k=1$) soldered to an SMA connector was introduced into the copper enclosure to provide an output coupler.

The frequency of the STO-ring-loaded cavity's TE$_{01\delta}$ mode was mechanically tuned (varying the height of the cavity's internal metal  ``ceiling'' suspended on a screw) to Pc:Ptp's  $|\textrm{X}\rangle\leftrightarrow|\textrm{Z}\rangle$ transition, \textit{i.e.}~$f_{\text{mode}} = f_\textrm{XZ} =1.4495$ GHz. This coupled the microwave photons resident in the TE$_{01\delta}$ mode to the spin polarization associated with said transition. 
The loaded quality factor ($Q_\text{L}$) of the TE$_{01\delta}$ mode was measured to be 3600 using a vector network analyzer (Agilent 8753C). Due to the high dielectric constant ($\varepsilon_r\approx312$) of STO at microwave frequencies,
the magnetic mode volume of TE$_{01\delta}$ was only $V_\text{mode}\sim0.3$~cm$^3$ leading to a high Purcell factor\cite{breeze2015enhanced} for magnetic-dipole transitions. The magnetic energy density and magnetic field vector of the TE$_{01\delta}$ mode were simulated using COMSOL Multiphysics with a 2D axisymmetric model\cite{oxborrow2007traceable}, and are shown in Fig.~\ref{fig.6}. Magnetic flux is ``funneled'' through the bore of the STO ring where the Pc:Ptp crystal is located, resulting in a single spin-photon coupling strength of $g_\textrm{s}=\gamma\sqrt{\mu_0hf_\text{mode}/2V_\text{mode}}=0.23$ Hz, where $\gamma$ is the electron gyromagnetic ratio, $\mu_0$ is the permeability of free-space, and $h$ is Planck's constant. Despite the weak coupling between a single spin and cavity photons, the spin-photon coupling strength can be significantly enhanced by a factor of $\sqrt{N}$\cite{kaluzny1983observation} when an ensemble of $N$ spins collectively interact with the photons within a small mode volume. This is why the Pc:Ptp system has also been proposed for quantum spin memories exploiting strong spin-photon coupling\cite{breeze2017room}. One can calculate, using the density of Ptp, that the number of pentacene molecules within an mm-sized 0.1\% doped Pc:Ptp crystal (such as in Fig.~\ref{fig.5}(c) with an approximate volume of 100 mm$^3$) can easily reach 10$^{17}$. A sufficiently large $N$, and thus factor of $\sqrt{N}$, can be generated if a sufficient fraction of these molecules can be photo-excited (spin-selectively) into T$_1$.

The magnetic-dipole moment associated with transitions between the $|\textrm{X}\rangle$ and $|\textrm{Z}\rangle$ sublevels, as is associated with the  $S_\textrm{Y}$ spin-operator, lies along pentacene's short in-plane molecular axis (\textit{i.e.}~its y-axis)\cite{yang2000zero}.
Given (i) the predominately axial polarization of the oscillating magnetic field that the  
TE$_{01\delta}$ mode provides, (ii) the fact that the grown crystal's predominate
cleavage plane contained this axis and (iii) the orientations of the two magnetically inequivalent
sites within Ptp's unit cell that Pc can dope into\cite{lang2005orientational,lang2007dynamics}, it is expected that the average dimensionless magnetic-dipole-transition matrix element (squared)\cite{siegmann1964microwave} for the  
stimulated emission (or absorption) between the $|\textrm{X}\rangle$ and $|\textrm{Z}\rangle$ sublevels will be a reasonably sizeable fraction of unity (depending on the exact crystal orientation).
Determining its value experimentally requires performing ``quantitative EPR'', which is complicated by uncertainty in orientation combined with the existence of inhomogeneous broadening, meaning that not all excited pentacene molecules have $|\textrm{X}\rangle\leftrightarrow|\textrm{Z}\rangle$ transitions (at zero field) in tune with the cavity's mode, namely $f_{\text{mode}} \neq f_\textrm{XZ}$. 

\section{Maser performance}
\begin{figure}
\centering\includegraphics{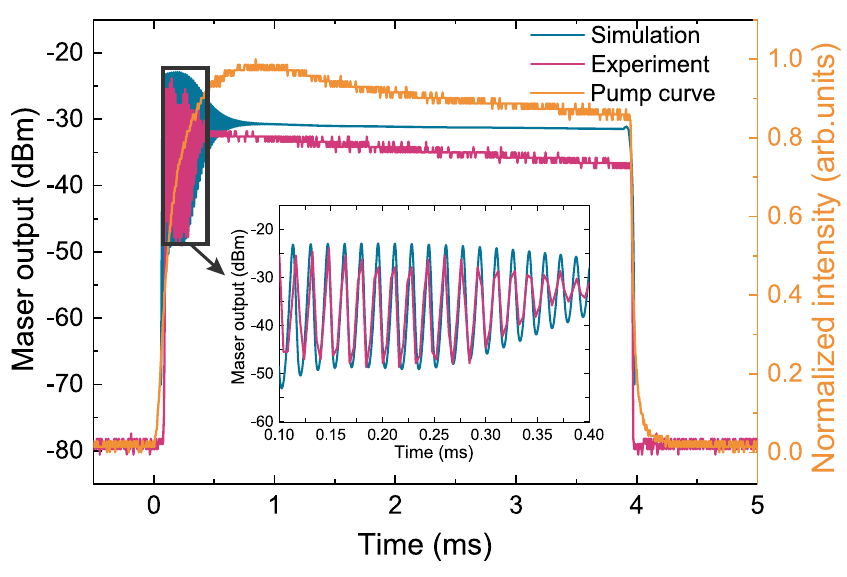}
\caption{Experimental (pink) and simulated (blue) 4-ms maser response evolving with the optical pumping of the Ce:YAG LC. Inset: Zoom-in of the Rabi oscillations occurring at the beginning of the maser burst.} \label{fig.7}
\end{figure}

A single pulse (or ``flash'') emanating from the xenon lamp inside the head of the IPL source lasts 4 ms and provides a fluence
 of 2.35 J/cm$^2$ (exiting the output face of the fused-silica light guide into air, measured by a Thorlabs ES245C pyroelectric sensor); see Fig.~\ref{fig.7}. 
The flash irradiates a rectangular region, 5~mm wide by 34~mm long (equating to an area of $\sim$1.7~cm$^2$) on one of the Ce:YAG LC's main faces, implying that the LC receives a $\sim$4 J dose of white pump light per flash. Since, the luminescent lifetime of the Ce:YAG is only $\sim$70~ns\cite{crytur17}, the instantaneous-power-versus-time profile of the LC's
yellow output faithfully mimics (\textit{i.e.}~same duration and shape as) that of the primary white pump flash.  
The energy of this secondary ``maser-pump'' pulse,
entering the Pc:Ptp crystal through the Ce:YAG LC's output faces, cannot be directly measured. However, 
adopting the reported power-conversion efficiency of $\sim$17\%\cite{sathian2017solid} for a similar Ce:YAG LC (with a mirror at one end), we estimate this pump energy to be 680 mJ corresponding to an instantaneous pump power
of $\sim$170~W. 
Upon the Pc:Ptp crystal receiving its pump pulse, 
a maser burst at 1.4495 GHz,  with a peak power of approximately -25 dBm,
could be detected using a logarithmic detector (Analog Devices AD8318) and was recorded using a digital storage oscilloscope (Tektronix TBS 1102B-EDU, 2 GSa/s sampling-rate, 100 MHz bandwidth); see Fig.~\ref{fig.7}. The maser burst also has virtually the same duration as that of its optical pump, namely 4~ms (excluding a few $\mu$s of start-up delay).

The oscillations observed at the beginning of the maser burst are Rabi oscillations resulting from coupling between the pentacene's spin ensemble and microwave photons in the TE$_{01\delta}$ mode. Such ``Dicke states''  have been studied for
a similar maser pumped with 5.5~ns pulses of 592-nm light, supplied by an OPO (pumped by a Q-switched Nd:YAG laser)\cite{breeze2017room}. This is a vastly shorter pump duration than that which is used in the present work (4~ms). 
Since the measured maser output $P_{\text{maser}}$ is related to the average number of photons $\langle\hat{a}^\dagger\hat{a}\rangle$ in the TE$_{01\delta}$ mode according to the expression $P_{\text{maser}} = hf_{\text{mode}}\langle\hat{a}^\dagger\hat{a}\rangle\kappa_\textrm{c}k/(1+k)$\cite{breeze2017room}, where $k=1$ is the coupling coefficient of the output port of the cavity, the observed 4-ms maser burst as well as the Rabi oscillations have been simulated (see Fig.~\ref{fig.7}) by extracting the average photon number $\langle\hat{a}^\dagger\hat{a}\rangle$ from the following master equations (Eq.~\ref{eq:3}-\ref{eq:8})\cite{jin2015proposal}:
\begin{widetext}
\begin{eqnarray}\label{eq:3}\frac{d\langle\hat{N}_\textrm{X}\rangle}{dt}&&=P_\textrm{X}\eta_\textrm{ISC}\mathit{\Gamma}-k_\textrm{X}\langle\hat{N}_\textrm{X}\rangle-\gamma_\textrm{XZ}(\langle\hat{N}_\textrm{X}\rangle-\langle\hat{N}_\textrm{Z}\rangle)-\gamma_\textrm{XY}(\langle\hat{N}_\textrm{X}\rangle-\langle\hat{N}_\textrm{Y}\rangle)-ig_\textrm{s}(\langle\hat{S}_+\hat{a}\rangle-\langle\hat{a}^\dagger\hat{S}_-\rangle)\\
 \label{eq:4}
    \frac{d\langle\hat{N}_\textrm{Y}\rangle}{dt}&&=P_\textrm{Y}\eta_\textrm{ISC}\mathit{\Gamma}-k_\textrm{Y}\langle\hat{N}_\textrm{Y}\rangle-\gamma_\textrm{YZ}(\langle\hat{N}_\textrm{Y}\rangle-\langle\hat{N}_\textrm{Z}\rangle)-\gamma_\textrm{XY}(\langle\hat{N}_\textrm{Y}\rangle-\langle\hat{N}_\textrm{X}\rangle)\\
\label{eq:5}
    \frac{d\langle\hat{N}_\textrm{Z}\rangle}{dt}&&=P_\textrm{Z}\eta_\textrm{ISC}\mathit{\Gamma}-k_\textrm{Z}\langle\hat{N}_\textrm{Z}\rangle-\gamma_\textrm{XZ}(\langle\hat{N}_\textrm{Z}\rangle-\langle\hat{N}_\textrm{X}\rangle)-\gamma_\textrm{YZ}(\langle\hat{N}_\textrm{Z}\rangle-\langle\hat{N}_\textrm{Y}\rangle)+ig_\textrm{s}(\langle\hat{S}_+\hat{a}\rangle-\langle\hat{a}^\dagger\hat{S}_-\rangle)\\
\label{eq:6}
    \frac{d\langle\hat{S}_+\hat{a}\rangle}{dt}&&=-\frac{\kappa_\textrm{c}+\kappa_\textrm{s}}{2}\langle\hat{S}_+\hat{a}\rangle-ig_\textrm{s}(\langle\hat{S}_+\hat{S}_-\rangle+\langle\hat{N}_\textrm{X}\rangle+\langle\hat{a}^\dagger\hat{a}\rangle\langle\hat{S}_\textrm{Z}\rangle)\\
\label{eq:7}
    \frac{d\langle\hat{S}_+\hat{S}_-\rangle}{dt}&&=-\kappa_\textrm{s}\langle\hat{S}_+\hat{S}_-\rangle+ig_\textrm{s}\langle\hat{S}_\textrm{Z}\rangle(\langle\hat{S}_+\hat{a}\rangle-\langle\hat{a}^\dagger\hat{S}_-\rangle)\\
\label{eq:8}
    \frac{d\langle\hat{a}^\dagger\hat{a}\rangle}{dt}&&=-\kappa_\textrm{c}(\langle\hat{a}^\dagger\hat{a}\rangle-\bar{n})+ig_\textrm{s}(\langle\hat{S}_+\hat{a}\rangle-\langle\hat{a}^\dagger\hat{S}_-\rangle)
\end{eqnarray}
\end{widetext}
 Here, the population of pentacene molecules in a state or sublevel $\textrm{L} \in \{ |\text{X}\rangle,\ |\text{Y}\rangle,\ |\text{Z}\rangle \}$ is expressed as the expectation value of the number operator $\hat{N}_\textrm{L}$, respectively. $\hat{S}_+\hat{a}$, $\hat{S}_+\hat{S}_-$ and $\hat{a}^\dagger\hat{a}$ are, respectively, the operators of spin-photon coherence, spin-spin coherence and the photon number in the TE$_{01\delta}$ mode. $\hat{S}_\textrm{Z} = \hat{N}_\textrm{X}-\hat{N}_\textrm{Z}$ is the spin polarization between the $|\text{X}\rangle$ and $|\text{Z}\rangle$ triplet sublevels.
 The optical pump rate is evaluated as
 $\mathit{\Gamma} = \eta_\textrm{opt} P_{\text{pump}}/hf_{\text{pump}}$,
 where $\eta_\textrm{opt}$ is a dimensionless parameter adjusting the overall efficiency of the pumping process (it has no relation to the previously mentioned maser ``co-operativity''). We found the best-fit value of the efficiency $\eta_\textrm{opt}$ is about 5.4\%. The inefficiency can be attributed to (i) the inhomogeneous broadening of pentacene ($\sim 4$ MHz\cite{oxborrow2012room}) such that only about 10\% of the pentacene molecules with the $|\text{X}\rangle \leftrightarrow|\text{Z}\rangle$ transition match the cavity's linewidth ($\sim$ 400 kHz), and (ii) the misalignment between the mode's magnetic field and pentacene's magnetic-dipole moment lying along the short in-plane molecular axis.
 $P_{\text{pump}}$ is the instantaneous power of the pump ``beam'' coming from the Ce:YAG luminescent concentrator;
 $f_{\text{pump}}$ is the optical frequency of the LC's peak emission at 528 nm; and 
 $\eta_\textrm{ISC}$ is the ISC yield ($\sim 62.5\%$\cite{takeda2002zero}) that can be evaluated according to $\eta_\textrm{ISC} = k_\textrm{ISC}/(k_\textrm{ISC}+k_\textrm{sp})$. The rate of the spontaneous emission $k_\textrm{sp} = 4.2\times10^7$ s$^{-1}$\cite{takeda2002zero} and the ISC rate $k_\textrm{ISC} = 6.9\times10^7$~s$^{-1}$\cite{takeda2002zero}.
 $P_\textrm{X} = 0.76, P_\textrm{Y} = 0.16$ and $P_\textrm{Z} = 0.08$ are the normalized ($P_\textrm{X} + P_\textrm{Y} + P_\textrm{Z} = 1$) spin-selective population rates for transfers of population mediated by inter-system crossing into 
 T$_1$'s $|\textrm{X}\rangle$, $|\textrm{Y}\rangle$ and $|\textrm{Z}\rangle$ sublevels. 
 Because internal conversion from T$_2$ to T$_1$ (in Fig.~\ref{fig.1}) is both extremely fast and spin-polarization preserving, one can regard the ISC process as moving population directly from S$_1$ into T$_1$'s
 three sublevels.
 
 Our best-fitted triplet decay rates ($k_\textrm{i}$) and the spin-lattice relaxation rates ($\gamma_\textrm{ij}$) are shown in Table.~\ref{tab:table1}.
\begin{table}[b]
\caption{\label{tab:table1}Zero-field spin dynamics of pentacene's lowest triplet state T$_1$. Units: 10$^4$ s$^{-1}$.}
\begin{ruledtabular}
\begin{tabular}{ccccccl}
$k_\textrm{X}$&$k_\textrm{Y}$&$k_\textrm{Z}$&$\gamma_\textrm{XZ}$&$\gamma_\textrm{YZ}$&$\gamma_\textrm{XY}$&\\
\colrule
2.8$\pm$0.5 & 0.6$\pm$0.2 & 0.2$\pm$0.09 & 1.1$\pm$0.2 & 2.2$\pm$0.2 & 0.4$\pm$0.2 & ref.~[\onlinecite{wu2019unraveling}]\\
2.2 & 1.4 & 0.2 & 1.1 & 2.8 & 0.4 & This work \\
\end{tabular}
\end{ruledtabular}
\end{table}
We note that the values obtained for $k_\textrm{X}$, $k_\textrm{Y}$ and $\gamma_\textrm{YZ}$ differ slightly from those reported in our previous study\cite{wu2019unraveling}.
 We suspect that the discrepancies stem from the different light sources used for pumping the Pc:Ptp crystal. In ref.~\cite{wu2019unraveling}, an optical parametric oscillator (OPO), providing a monochromatic output at 590 nm, was used. Light at this wavelength can only be absorbed via pentacene's S$_0\rightarrow$S$_1$ transition.
 In contrast, the Ce:YAG LC's output spans a relatively broad range of wavelengths as shown in Fig.~\ref{fig.4}, which may further induce the optical excitation of pentacene molecules in T$_1$ to higher triplets states
 (\textit{e.g.}~$\textrm{T}_1\rightarrow \textrm{T}_3$\cite{hellner1972absorption,khan2017theory,rao2011photophysics}), potentially altering the triplet spin dynamics\cite{iinuma1997dynamic}. $\kappa_\textrm{c} = 2\pi f_\textrm{XZ}/Q_{\text{L}} = 2.5$ MHz is the cavity decay rate. $\kappa_\textrm{s} = 1.1$ MHz is the spin-dephasing rate of the pentacene's triplets which is larger than the value measured by zero-field pulsed EPR free induction decay (FID) in a separate study (0.7~MHz)\cite{yang2000zero}. The difference may arise from a concentration effect\cite{morton2007environmental}: our Pc:Ptp crystal had a doping concentration of 0.1\% whereas that of the crystal studied in the  FID measurements was lower at $\sim$ 0.01\%. 
 $\bar{n}=1/(\text{exp}[hf_{\text{mode}}/k_{\text{B}}T]-1)\approx4300$ is the mean number of thermal photons in the TE$_{01\delta}$ mode where $k_{\text{B}}$ is the Boltzmann constant and $T=298$ K is the temperature. The coupled master equations were integrated in time with initial conditions 
 $\langle\hat{N}_\textrm{X}\rangle = \langle\hat{N}_\textrm{Y}\rangle = \langle\hat{N}_\textrm{Z}\rangle = \langle\hat{S}_+\hat{a}\rangle = \langle\hat{S}_+\hat{S}_-\rangle = 0$ and $\langle\hat{a}^\dagger\hat{a}\rangle = \bar{n} \approx4300$.

As revealed in Fig.~\ref{fig.7}, the simulation reproduces the Rabi oscillations alongside the 4-ms maser burst. The experimental maser oscillation exhibits a decrease in output power over the 4-ms duration while the simulated output power is stable. We attribute the discrepancy to thermal-loading of the gain medium. By integrating the maser signal, the output energy can be estimated to be $\sim$1.55 nJ and since the pentacene molecules have low fluorescence quantum yield ($\sim$27.5\%\cite{takeda2002zero}) and low phosphorescence yield (due to dominant non-radiative triplet decay\cite{clarke1976triplet}), the optical pump energy (680 mJ) can be assumed to be mostly converted to heat. The heat capacity of the host material \textit{para}-terphenyl is $\sim$ 280 J K$^{-1}$ mol$^{-1}$ at room-temperature\cite{chang1983heat} and the Pc:Ptp crystal contains around $4.4\times10^{-4}$ mol of \textit{para}-terphenyl. Therefore, a single 4-ms optical pump pulse should raise the temperature of the crystal by $\sim$6 K. The temperature increase will give rise to faster spin relaxation processes in pentacene's triplet sublevels\cite{ong1995deuteration}
and shifts in both the frequency of  the maser transition  $|\textrm{X}\rangle\leftrightarrow|\textrm{Z}\rangle$
and the frequency of maser cavity's  TE$_{01\delta}$ mode (upon the crystal transferring its acquired warmth to the STO surrounding it).
Using Figure 4 of ref.~\cite{lang2007dynamics}, the temperature dependence of the former can be
estimated to be -80~kHz/K. Considering an extreme case that the optical pump energy is completely transferred to the dielectric ring containing $2.3\times10^{-2}$ mol of STO whose heat capacity is $\sim100$ J K$^{-1}$ mol$^{-1}$\cite{de1996high}, the maximum temperature rise of the STO should be $\sim 0.3$ K.
Using the known Curie-Weiss temperature dependence of STO's dielectric constant \cite{saifi1970dielectric}, the temperature dependence of the latter is estimated to be around +2.6~MHz/K. These changes could certainly degrade the maser's output power as seen in Fig.~\ref{fig.7}. The observed thermal issue implies that, as with many types of CW lasers, active cooling of the gain medium will be required for sustained operation and that, if the resonator is adjusted to match the frequency of the maser transition when the crystal is continuously pumped (and thus warm),  there could be an initial delay in the onset of masing as the crystal warms up.   
\begin{figure}
    \centering
    \includegraphics{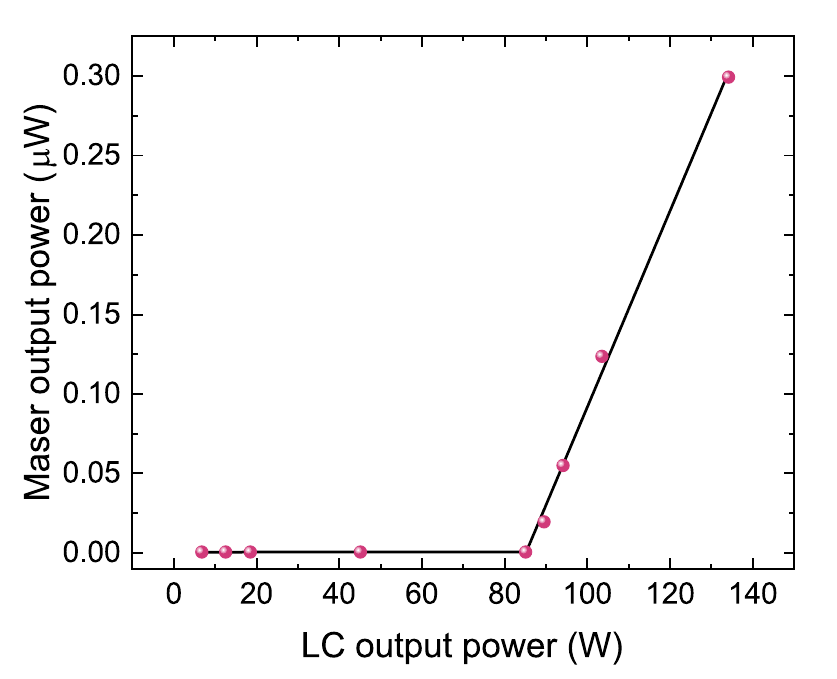}
    \caption{Determination of the optical pump threshold of the 4-ms pentacene maser.}
    \label{fig:threshold}
\end{figure}

The previous study\cite{oxborrow2017maser} showed that a certain figure of merit, $\kappa$, with an expression of
\begin{eqnarray}
    \kappa =&& P_\textrm{X}[1+(k_\textrm{Y}+k_\textrm{Z}+k_\textrm{Y}k_\textrm{Z}/\gamma_\textrm{YZ})/k_\textrm{X}]\nonumber\\
    &&-(P_\textrm{Z}/\gamma_\textrm{YZ})\{k_\textrm{Y}+\gamma_\textrm{XY}[1+(k_\textrm{Y}+k_\textrm{Z})/k_\textrm{X}]\}\nonumber\\
    &&+(k_\textrm{Z}/k_\textrm{X})(\gamma_\textrm{XY}/\gamma_\textrm{YZ})-1
\end{eqnarray}
must exceed zero for continuous masing to be possible. Using the spin dynamics extracted from the simulation, we evaluated that $\kappa = 0.3$, which indicates that our setup in this work is feasible for operating a pentacene maser in a CW mode at room-temperature.

In addition, the optical pump threshold of the 4-ms pentacene maser was determined to be 85 W via the pump power dependent measurements, as shown in Fig.~\ref{fig:threshold}.

\section{Conclusion and outlook}
In summary, pumped by a Ce:YAG LC with an output power of 170 W, a room-temperature quasi-CW pentacene maser has been successfully achieved. It can operate continuously for 4 ms with a peak power of -25 dBm. The optical pump threshold of the 4-ms maser was measured to be 85 W. Although this is preliminary work for realizing a CW pentacene maser, the results have solved a long-standing mystery about whether the pentacene maser can work in a CW mode or not. We found that it is perfectly possible for the pentacene maser to run in a CW mode provided the  heat generated by optical pumping of the gain medium can be removed  effectively. Flowing argon gas around the pentacene gain medium might be a solution. Alternatively, efforts can be made to explore a maser crystal which simultaneously possesses the attributes of high spin polarization density (analogous to the Pc:Ptp system), slow spin-lattice relaxation and robust thermal stability at room temperature. Moreover, similar to LC pumped CW lasers\cite{yang2009development,barbet2016light}, a high power light-emitting diode could be used, in lieu of a flash lamp, to pump the Ce:YAG LC (combined with invasive optical pumping) and power the pentacene maser. As an efficient pump source, LCs offer the advantages of low cost, low maintenance and simple fabrication, which would support the future development of cheaper, high-power solid state lasers/masers. 
\section*{Funding}
This work was supported by the U.K. Engineering and Physical Sciences Research Council through grants EP/K037390/1 and EP/M020398/1. H.W. acknowledges financial support from the China Scholarship Council (CSC) and Imperial College London for a CSC-Imperial PhD scholarship 201700260049.

\begin{acknowledgments}
We thank Ben Gaskell of Gaskell Quartz Ltd.~for hand-fabricating (with no little care and precision) the strontium titanate ring used. We thank  Simon Rondle of Photronics Ltd.~for supplying, refurbishing  and commissioning  the xenon flash lamp. We also thank Dr. Shamil Mirkhanov for the preliminary work on invasive optical pumping.

\end{acknowledgments}

%


\end{document}